# Dynamics of single vortices in grain boundaries: I-V characteristics on the femto-volt scale

B. Kalisky[1,*], J. R. Kirtley[1], E. A. Nowadnick[1], R. B. Dinner[1,a], E. Zeldov[1,2], Ariando[3,b], S. Wenderich[3], H. Hilgenkamp[3], D. M. Feldmann[4], K. A. Moler[1]

1. Geballe Laboratory for Advanced Materials, Stanford University, Stanford, CA 94305
2. Department of Condensed Matter Physics, Weizmann Institute of Science, Rehovot 76100, Israel
3. Low Temperature Division, Mesa+ Institute for Nanotechnology, University of Twente, P. O. Box 217, 7500 AE Enschede, The Netherlands
4. Los Alamos National Laboratory, Los Alamos, New Mexico 87545, USA

We employed a scanning Hall probe microscope to detect the hopping of individual vortices between pinning sites along grain boundaries in $YBa_2Cu_3O_{6+\delta}$ thin films in the presence of an applied current. Detecting the motion of individual vortices allowed us to probe the current-voltage (I-V) characteristics of the grain boundary with voltage sensitivity below a femto-volt. We find a very sharp onset of dissipation with $V \propto I^n$ with an unprecedented high exponent of $n \approx 290$ that shows essentially no dependence on temperature or grain boundary angle. Our data have no straightforward explanation within the existing grain boundary transport models.

Grain boundaries (GB) in high-$T_C$ superconductors[1] are widely used for devices and fundamental studies[2,3]. They also govern the transport properties of many superconductors that have potential technological applications[4]. It is therefore of great importance to understand the mechanism for dissipation in transport across GB's. GB transport is usually studied with a voltage threshold of nanovolts or higher. In this regime, millions of vortices traverse the boundary per second. Although not on a grain boundary, effective voltage thresholds of picovolts have been reached using a SQUID voltmeter[5], and noise spectral densities of milliattovolts have been reported by measuring flux noise in superconducting rings[6]. Extremely small effective voltages have been inferred from measurements of flux relaxation in superconducting rings[7,8]. Here, we report direct imaging of grain boundaries in $YBa_2Cu_3O_{6+\delta}$ with a scanning Hall probe microscope[9]. By sensing individual vortices, we demonstrate an effective voltage threshold of 0.2 femtovolts and quantitatively measure dissipation with an onset threshold of attowatts. The results are striking: the frequency of vortex motion increases exponentially with current above a threshold current, but the exponent does not depend strongly on either temperature or grain boundary misorientation angle. We compare our data to general models of vortex motion in grain boundaries[1,10-16], and surprisingly find that none of the existing models can fully describe the data.

The data shown here are from a thin film of $YBa_2Cu_3O_{6+\delta}$ ($T_C$=89.5K), which was epitaxially grown by pulsed laser deposition on a substrate of symmetric 24° [001] tilt oriented bicrystal of $SrTiO_3$. The film, 250 nm thick, was patterned into 21 bridges with widths of 20, 50 and 100 μm, so that the GB was perpendicular to the bridge (see Fig.

1a). We also measured a 50 μm wide bridge patterned from a 180nm thick film grown on a 5° [001] bicrystal. Magnetic measurements were performed with a custom-built large-area scanning Hall probe microscope[9], with single vortex resolution and a 10 kHz bandwidth. The magnetic resolution of the system is 4 mG/√Hz and the spatial resolution in the data presented is ≥1μm (the scanner pixel size is 200nm, 0.5-2 μm sized Hall bars were used, and the data shown were taken with a 1 μm x 1 μm Hall bar with an effective sample surface to sensor distance of 1 μm). The Hall probe is moved over the sample surface, pausing at each location as the Hall voltage is recorded. Using the same setup, 4-probe transport measurements of the sample were made with a voltage noise floor of 1 μV.

Large angle GB's with critical currents much lower than the surrounding grains[1,12,14] provide a well defined narrow channel for vortex motion. Figure 1b shows a magnetic image of a 50 μm wide GB, at 10 K, with applied current of I=29 mA, larger than the critical current of the GB. The amplitude of the fields due to the edge currents is ~5 G, and single vortices are observed in the bulk due to cooling in an ambient field of ~0.15 G. These bulk vortices are well pinned at this applied current. On the GB, however, vortices are continuously moving from both edges towards the central (white) region where positive and negative vortices annihilate. In order to estimate the number and position of vortices moving on the GB, we fit this image using the monopole model for each single vortex, with the position of each vortex as a free parameter. The fit gives a total of 10 negative (blue) and 23 positive (red) vortices on the 50 μm GB, which are unevenly spaced with a minimum distance of 1 μm between neighboring vortices. When the applied current is turned off, many vortices stay trapped on the GB. However, after thermal cycling to 70 K, only a few of them are left and we can image single vortices on the GB after cooling back to 10 K as shown in Fig. 1c. Vortices on the GB do not look different from bulk vortices with our spatial resolution, indicating $\lambda_{GB} \leq 1\mu m$.

We now describe the vortex dynamics along the GB as measured by the Hall probe in the time and frequency domains. Examples for the time dependence of the Hall probe signal with time are shown in Fig. 1d-1e. Figure 2 shows a pseudocolor image of the noise spectrum for different applied currents, measured at 10K at location #1 indicated in Fig. 1b. A large enhancement of the low-frequency noise (color-coded red-yellow in Fig. 2) appears when the applied current exceeds a temperature dependent threshold value $I_{th}$. The spectrum broadens rapidly with further current increase. Similar behavior was observed at different locations along the GB but disappeared in the grains 2-3 μm away from the GB. The appearance of the excess noise only on the GB location and its dependence on the sample current relates it to the dynamics of vortices which are confined to the 1D channel of the GB.

The noise spectrum has a Lorentzian shape (inset of Fig. 2) consistent with a two-level random telegraph noise [17,18], implying a random hopping of vortices along the GB, rather than a smooth correlated motion. The shift of the Lorentzian to higher frequencies with increasing current shows that the vortices hop faster with increasing Lorentz force as expected. At each current above the threshold, the total value of the excess noise (difference between the integrals of the power spectral density above and below $I_{th}$) remains rather constant and matches the peak value measured for a single vortex by the static imaging (~1 G).

The motion of vortices is also apparent in the time resolved Hall probe signal. Figures 1d and 1e show the time traces at various applied currents at two locations along the GB marked in Fig. 1b. The signal was recorded for 16 s ($3 \times 10^5$ data points) at each applied current, and a 60 Hz background noise (see Fig. 2) was filtered out from the raw data. For I < 28.5 mA no vortex motion was detected as seen in the corresponding trace in Fig. 1d. At 28.5 mA individual rare switching events become visible, with the amplitude of a single vortex. The rate of switching events increases with increasing current. Above a certain current the vortex rate exceeds our bandwidth. No correlation was observed between events, which points again to the scenario of random vortex hopping between pinning sites. At different locations along the GB, different types of signals were observed – positive vortices (in the red region), negative vortices (blue region), and a mixture of the two close to the annihilation region. Also, in some cases 3 or 4 level signals were observed at certain locations along the GB, which can be explained by the presence of a few pinning sites in close proximity to the Hall sensor. An example of a 3 level signal is shown in Fig 1e, recorded at 10 K at location #2 in Fig. 1b (1 um away from location #1). Such signals were more frequently observed at the higher temperature of 50 K, possibly because vortices are larger and overlap more. The fact that at this site, the highest level always precedes the intermediate level indicates that the vortices are moving in one direction as opposed to hopping back and forth.

From these data the current-voltage (I-V) characteristics were extracted, using the Josephson relation $V = \Phi_0 f$, where $\Phi_0$ is the flux quantum. The rate of passing vortices, f, was measured either by counting the vortices that appear on the time traces (e.g. Fig. 1d-1e) or by fitting the noise spectrum (Fig. 2) to a Lorentzian describing two-level random telegraph noise. Figure 3 shows I-V curves at 10 K (red) and 50 K (blue). The current axis is normalized by $I_{th}(10K)=28.5$ mA, which is the lowest current at which vortices were observed moving at 10 K. Such "magnetic" measurements of the voltage on a single vortex scale allow detection of voltages down to 0.2 femtovolts, where the average rate of passing vortices is ~0.1 Hz. The onset of dissipation is quite steep: at 10 K the voltage scale changes by 3 orders of magnitude as the applied current changes by $\delta I/I = 0.03$, and at 50K $\delta I/I=0.037$. The low voltage part of I-V can be equally well fitted by power law $V \propto I^n$ or exponential behavior $V \propto \exp(mI/I_c)$, where m, n, $I_c$ are fitting parameters. For both, the characteristic exponent is very high, $n \approx m \approx 290$. This characteristic exponent appears to be independent of temperature in the measured range of 10-50 K. For the 5° grain boundary at 40K we find $\delta I/I=0.04$. In all of the measurements we made $\delta I/I$ was between 0.03 and 0.05. Note that although the exponent is not temperature dependent, the critical current, at which dissipation starts, decreases with temperature as expected[10,14] since the gap and order parameter are decreasing. At higher voltages, conventional 4-probe transport measurements are possible, as shown by the green data points[19]. The shape of the high voltage part is similar to I-V curves observed in the literature [1,12,14]. For three of our 24° degree GB we found $I_cR_n$ products of 0.93, 1.68 and 0.94 mV at 10K, and our 5° GBJ had an $I_cR_n$ product of 0.4mV at 40K. At 86K the current-voltage characteristics had little hysteresis, consistent with our estimate of $\beta_c = 2\pi I_c R_n^2 C / \Phi_0 < 1$.

The Hall probe may perturb the sample, but we believe the possible perturbations are not important. The magnetic field generated by the Hall probe at the surface of the superconducting film is about 0.03G, which should be too small to affect the vortex dynamics. The tip of the Hall probe chip touching the sample may change the local superconducting properties through either the applied force (~1 μN) or the local heating (~10 μW).[9] The onset current for vortex motion changes slightly with the Hall probe's distance from the grain boundary, consistent with a local temperature rise of ~0.2K, but the shape and quantitative slope of the I-V curve remain the same. The amplitude of the magnetic signal shows a variation as a function of position perpendicular to the GB consistent with the shape of a single vortex, which is the expected dependence because we are scanning across a 1D channel of moving vortices. The noise data is consistent with the 4-probe IV measurements that were taken without the presence of the probe, in the sense that a relatively smooth curve can be drawn between the two sets of data (see Fig. 3).

Several scenarios were considered in order to explain the experimental observations:

(i) The shape of the I-V curves, the range of temperatures and the observed hopping of the vortices, which demonstrate the importance of pinning sites, suggest a *creep* behavior. Several pinning potential shapes have been considered [20,21] and although some of them predict an I-V shape that resembles the data, all of them also predict a temperature dependent exponent and therefore cannot describe our case. When relating our data to creep behavior we also consider the large "n number" obtained when fitting the curves to a power-law ($V \sim I^n$). The large n observed (~290) implies within the Bean model that the barrier is much larger than thermal fluctuations in the entire range 10-50K. Such a large n has not been reported in the literature.

(ii) For our 24 degree GB, Josephson vortices or *Abrikosov-Josephson vortices* [13,18,20] are expected. For the latter, a *non-linear flux flow* behavior is expected, due to the expansion of the vortex Josephson-core from $\xi$ to $\lambda$ as the current is increased [22]. Our measurements are performed in a range where the presence of Abrikosov-Josephson vortices is very likely[15,22]. However, this scenario still cannot explain the observed lack of temperature dependence of the exponent.

(iii) The presence of pinning, the steep curvature of the I-V and its decrease at higher voltages suggest a *glassy behavior*[20,23] with exponential I-V which can describe our data. However, care should be taken in relating this model to our case, which deals with 1D motion of vortices. Our vortices are 250 nm long and are relatively sparse, with a minimum distance of 1 µm between them. Also, the lack of temperature dependence of the exponent still remains a problem.

(iv) The lack of temperature dependence, and the very large pinning barrier relative to thermal fluctuations lead to considering the possibility of *quantum tunneling* of vortices between pinning sites. We have made an estimate of the quantum tunneling rate of Josephson vortices in our GBJ geometry. This estimate can be made to fit our experiments (see supplementary material). However, unrealistically high values for the attempt frequency and low values for the junction capacitance are required for these fits.

In summary, the dynamics of vortices on grain boundaries were measured for the first time in the limit of extremely low dissipation using scanning Hall probe microscopy. The vortices are observed to hop between pinning sites with a characteristic frequency that depends on the applied current. I-V characteristics were extracted from the data, with voltage sensitivity as good as $2 \times 10^{-16}$ V. At low voltages the I-V curves are very steep with temperature independent exponents. Our data are qualitatively different from grain-boundary transport measurements at higher voltages [1,12,14], and cannot be fully explained using existing models.

The authors would like to thank Janice Guikema and Clifford Hicks for Hall sensors, H. Karapetyan for help in measurements, and M. Beasley and A. Gurevich for useful discussions. This work is funded by an Air Force Multi-University Research Initiative (MURI), by the Center for Probing the Nanoscale (CPN), an NSF NSEC, NSF grant no. PHY-0425897, and by the US-Israel Binational Science Foundation (BSF). EZ acknowledges the support of EU FP7 ERC-AdG.

Figure Captions:

Figure 1: (a) Sketch of the grain boundary (GB) sample. The bridge width is 50 μm, GB misorientation angle is 24° symmetric (a-b axes are -12° and 12° with respect to the GB). (b,c) Magnetic image of the GB area (b) at 10 K and I=29 mA and (c) at 10 K after thermal cycling to 70K with I=0, showing individual vortices (red) and antivortices (blue). The regions imaged are marked by the orange frames in (a). (d,e) The Hall probe signals vs. time at 10 K for different currents applied to the GB measured at two positions marked in Fig. 1b. The switching events are vortices passing under the probe. Red circles indicate three-level switching events.

Figure 2: Pseudocolor image of the noise spectrum of the Hall signal for different sample currents at 10 K, measured at the location #1 in Fig. 1b. Inset: Difference between the power spectral densities at currents of 29 mA and 28mA (green). The Lorentzian shaped spectrum is multiplied by f similarly to the analysis in ref [16]. The fit to two-level random telegraph noise is plotted in red.

Figure 3: Current-voltage characteristics. The low voltage part (red, blue) was measured magnetically by the Hall probe (as shown in figs 1d and 2). Different symbols are different locations along the GB. The current axis is normalized by $I_{th}$(10K) which is the lowest current at which vortices were observed moving at 10K. The high voltage part (green) was acquired by four probe measurements on a neighboring bridge with identical dimensions and normalized to $I_{th}$ of that bridge measured magnetically (green data on the low voltage part). Inset: low voltage data at 10 K on a linear scale.

Current address:
* Email: beena@stanford.edu
*a. Department of Materials Science, Cambridge University, United Kingdom*
*b. Department of Physics, National University of Singapore, Singapore*

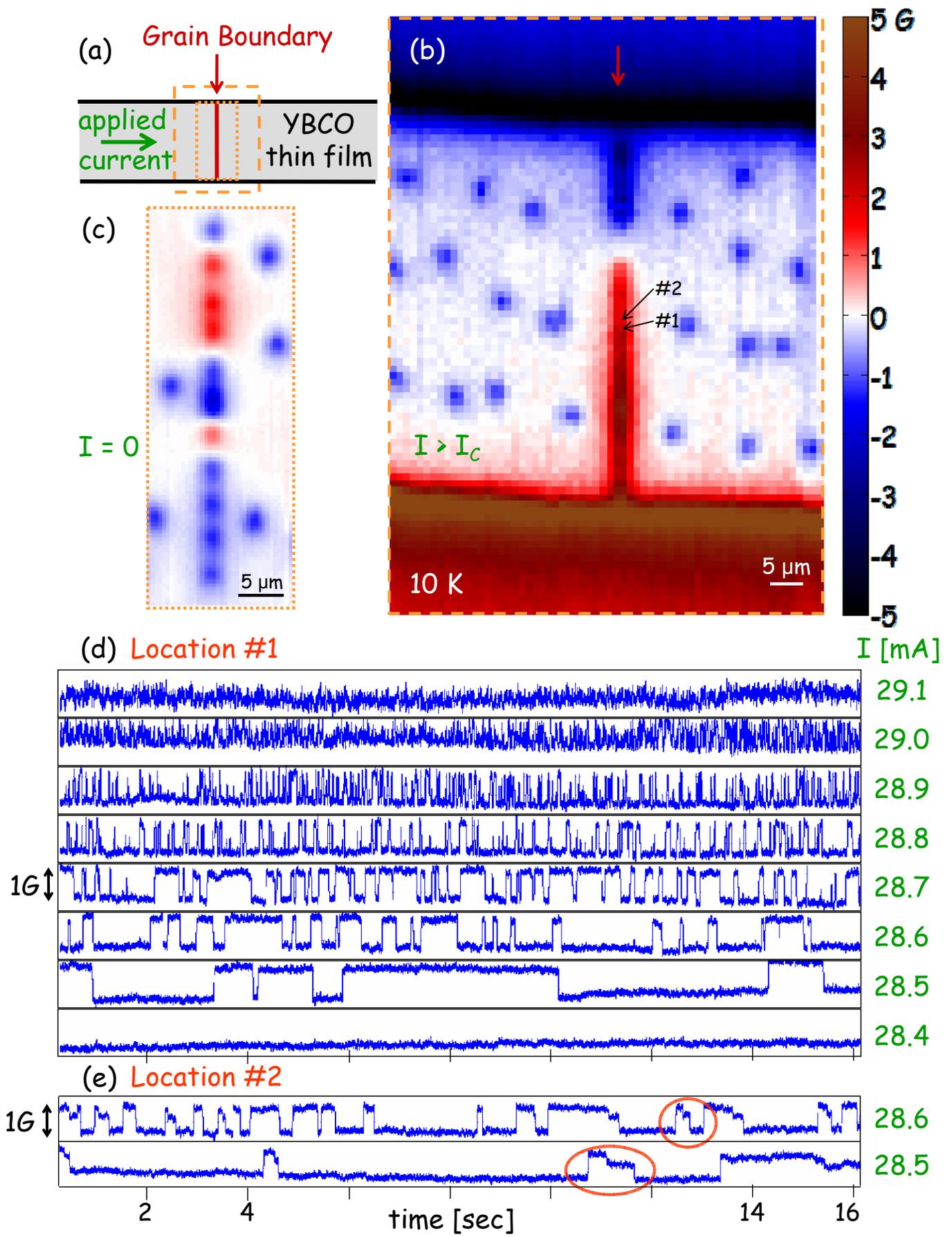

Fig. 1

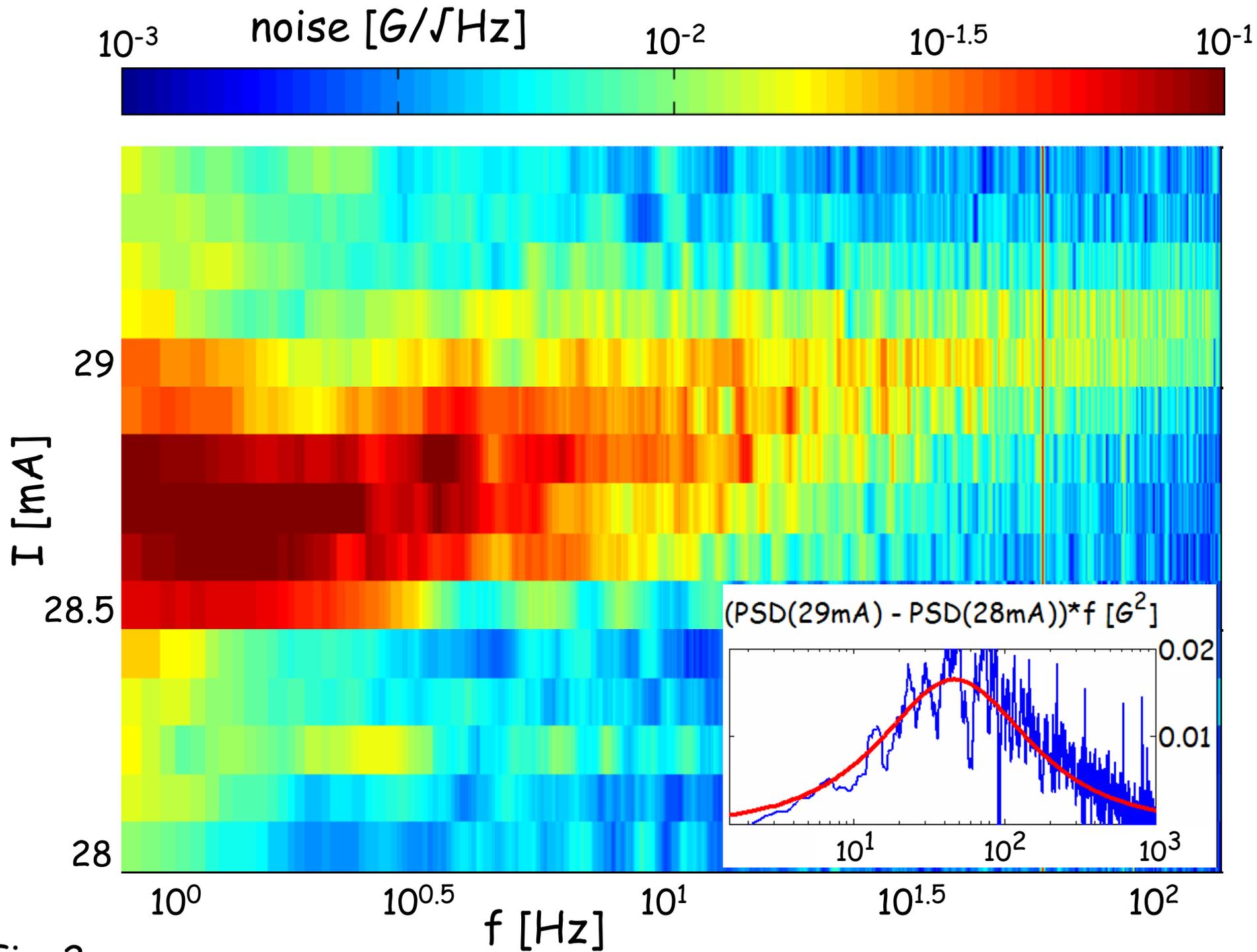

Fig. 2

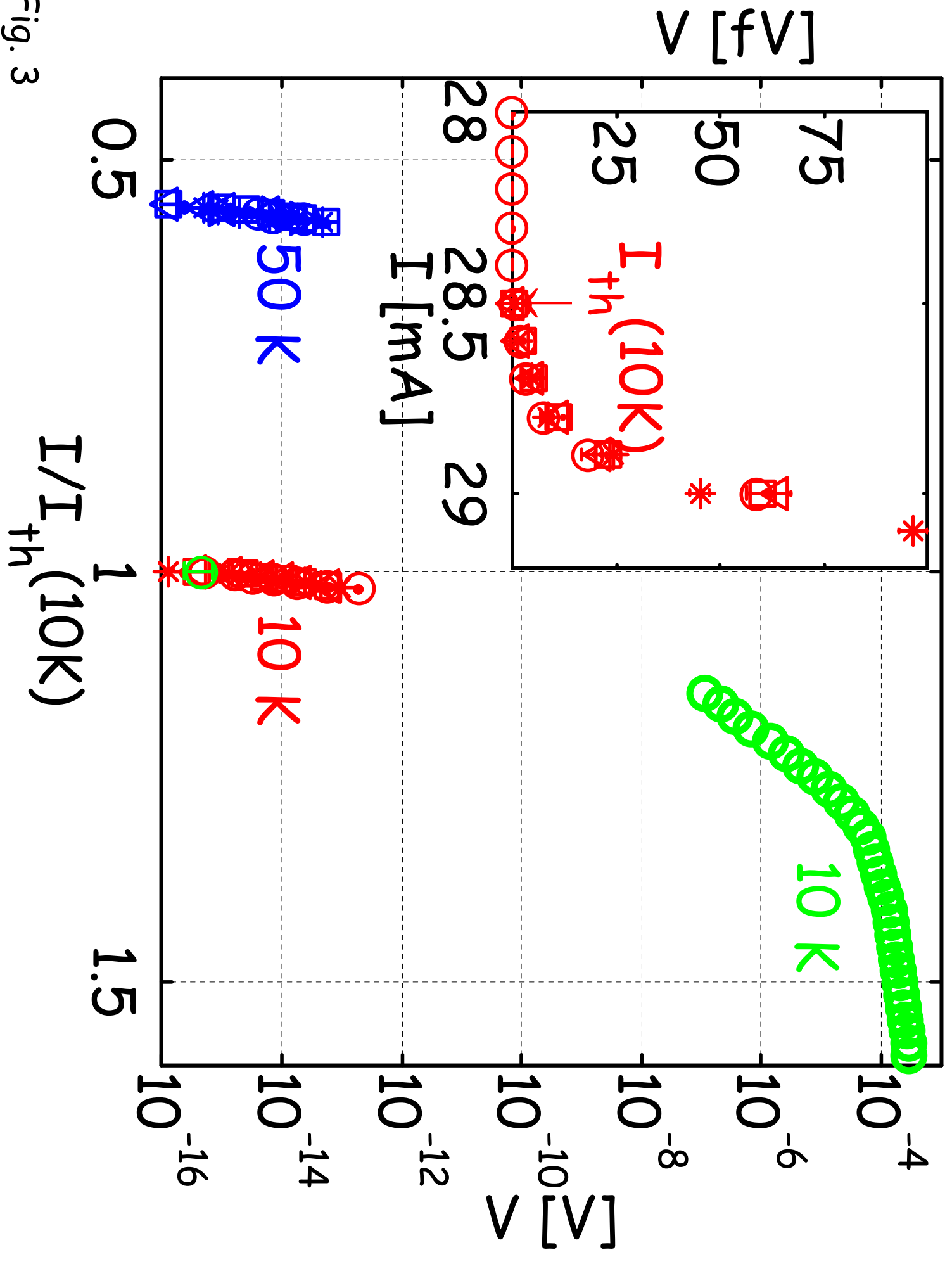

Fig. 3

Supplementary Material for:

Dynamics of single vortices in grain boundaries: I-V characteristics on the femto-volt scale

B. Kalisky[1,*], J. R. Kirtley[1], E. A. Nowadnick[1], R. B. Dinner[1,a], E. Zeldov[1,2], Ariando[3,b], S. Wenderich[3], H. Hilgenkamp[3], D. M. Feldmann[4], K. A. Moler[1]

1. Geballe Laboratory for Advanced Materials, Stanford University, Stanford, CA 94305
2. Department of Condensed Matter Physics, Weizmann Institute of Science, Rehovot 76100, Israel
3. Low Temperature Division, Mesa+ Institute for Nanotechnology, University of Twente, P. O. Box 217, 7500 AE Enschede, The Netherlands
4. Los Alamos National Laboratory, Los Alamos, New Mexico 87545, USA

**Quantum tunneling of Josephson vortices**

The free energy of a Josephson vortex in an infinitely long junction with no intrinsic phase shifts can be written as:[i]

$$(S.1) \quad F = \frac{hw}{2e}\int_{-\infty}^{\infty} dx\, j_c(x)[1-\cos(\phi(x)) + \frac{\lambda_J(x)^2}{2}\left(\frac{\partial \phi(x)}{\partial x}\right)^2].$$

Here $j_c(x)$ is the critical current density of the junction, $w$ is the film thickness, $\lambda_J(x) = (h/2\mu_0 e d j_c(x))^{1/2}$ is the Josephson penetration depth, $d$ is the spacing between the superconductors in the junction region, and $\phi(x)$ is the phase across the junction. The solution to the Sine-Gordon equation for the phase of a single Josephson vortex at $x=0$ in an infinite junction with constant $\lambda_J$ is given by:[i]

$$(S.2) \quad \phi(x) = \begin{cases} 2\pi - 4\tan^{-1}(e^{-x/\lambda_J}) & x > 0 \\ 4\tan^{-1}(e^{x/\lambda_J}) & x < 0 \end{cases}$$

The free energy of a Josephson vortex centered in a junction of width $L \gg \lambda_J$ with critical current $I_0$ is $F = 4\lambda_J \Phi_0 I_0 / \pi L$. Assume there is an inhomogeneity in the junction at $x = x_p$ over a length $\Delta x_p \ll \lambda_J$ with a local Josephson penetration depth $\lambda_{Jp}$ different from the average penetration depth $\overline{\lambda_J}$. For sufficiently short $\Delta x_p$ the inhomogeneity does not change $\phi(x)$ appreciably and the change in the free energy of the Josephson vortex is approximated by

$$(S.3) \quad \delta F = \frac{I_0 \Phi_0 \Delta x_p}{4\pi L}\left(\frac{\lambda_{Jp}^2}{\overline{\lambda_J}^2} - 1\right)\frac{4}{\cosh^2(x^*)} \,,$$

with $x^* \equiv x_p / \bar{\lambda}_J$. In the absence of an applied supercurrent, if $\lambda_{Jp} > \bar{\lambda}_J$ the inhomogeneity will produce a peak in the potential energy centered at $x = x_p$ with width $\Delta x = 2\bar{\lambda}_J \cosh^{-1}(\sqrt{2})$. There will be a dip in the potential energy for $\lambda_{Jp} < \bar{\lambda}_J$. It appears from our magnetic imaging that the supercurrent density across the junction is approximately independent of $x$ under conditions in which telegraph noise is observed. Then the potential energy associated with the Lorentz force on the vortex can be written as $F_L = I\Phi_0 x/L$, and the total free energy will have a peak or dip of the form given by Eq. S.3 added to a linearly sloping background from the Lorentz potential. The action for Josephson vortex tunneling through the potential barrier can then be written as

(S.4) $$S = \bar{\lambda}_J^{3/2} \left( \frac{m^* I_0 \Phi_0}{2\pi L} \right) \int_{x^*_{min}}^{x^*_c} dx^* \left[ \tilde{F} - \tilde{F}_{min} \right]^{1/2},$$

where $m^*$ is the effective mass of the Josephson vortex, the integral is taken over $x^*$ from $x^*_{min}$, the point where the dimensionless Josephson vortex potential $\tilde{F}$ has a local minimum $\tilde{F}_{min}$, to $x^*_c$, the next point at which $\tilde{F}$ becomes less than $\tilde{F}_{min}$, and

(S.5) $$\tilde{F} = 8 + \left( \frac{\lambda_{Jp}}{\bar{\lambda}_J^2} - 1 \right) \frac{4}{\cosh^2(x^*)} \frac{\Delta x_p}{2\bar{\lambda}_J} + \frac{2\pi I x^*}{I_0}.$$

The effective mass of the Josephson vortex can be written as[ii] $m^* = 8m/\beta^2$, where $m = \hbar/\bar{\lambda}_J^2 \omega_J$, $\omega_J = (2\pi I_0 / \Phi_0 C)^{1/2}$ is the Josephson plasma frequency, $C$ is the capacitance of the junction, and $\beta^2 = \frac{4e^2}{\hbar} \sqrt{\frac{\mu_0}{\varepsilon_0}} \frac{\sqrt{(2\lambda_L + d)d}}{w}$, where $\lambda_L$ is the London penetration depth of the superconductors making up the junction. In the limit of small dissipation, the quantum tunneling rate is given by

(S.6) $$\nu = \nu_0 e^{-S/\hbar},$$

where $\nu_0$ is an attempt frequency. The inclusion of dissipation is expected to reduce the tunneling rate, so using Eq. S.6 will, if anything, overestimate the quantum tunneling contribution to the observed telegraph noise frequencies.

The action predicted by Eq. S.4 decreases roughly linearly with applied current over a broad range of currents and parameters, so that the quantum tunneling contribution to the telegraph noise frequency is expected to increase exponentially with applied current through the junction, in agreement with experiment. However to determine whether quantum tunneling is in fact the mechanism for our observed noise, we must evaluate the expressions above numerically. We use the attempt frequency $\nu_0$, the junction critical current $I_0$, the junction capacitance $C$, the inhomogeneity width $\Delta x_p / \bar{\lambda}_J$, and the ratio of the Josephson penetration depths squared $\lambda_{Jp}^2 / \bar{\lambda}_J^2$ as fitting parameters. In addition, we take the London penetration depth $\lambda_L=0.15$ μm, the junction length L=50 μm, the film thickness w=0.2 μm, and the insulator thickness $d = 10$ nm. The best fit of the above model to our symmetric 24°, T=10K data is shown in Fig. S1. The best fit parameters are $\nu_0=1.04\text{e}40$ Hz$^{-1}$, $I_0$=0.035 A, $C$=1.72e-17 F, $\Delta x_p / \bar{\lambda}_J = 0.1$, $\lambda_{Jp}^2 / \bar{\lambda}_J^2 = 50.5$. We can determine the uncertainties in the various parameters by varying each in turn while optimizing the other 4, and finding the values for which the $\chi^2$ is doubled from the optimal value. This results in the limits $10^{25}$ sec$^{-1}< \nu_0 < 10^{80}$ sec$^{-1}$, $4\times10^{-17}$ F $< C < 7\times10^{-18}$ F, 0.031 A $< I_0 <$ 0.042 A, $0.09 < \Delta x_p / \bar{\lambda}_J < 0.68$, and $47.1 < \lambda_{Jp}^2 / \bar{\lambda}_J^2 < 350$.

The allowed fit values for the attempt frequency are at least 12 orders of magnitude too high, given that the Josephson plasma frequency is about $10^{12}$ sec$^{-1}$. Similarly, we estimate the capacitance of our junctions to be about 50 pF [iii], some 6 orders of magnitude larger than our fit values. We conclude that it requires physically unrealistic values to fit our experimental data to a quantum tunneling model.

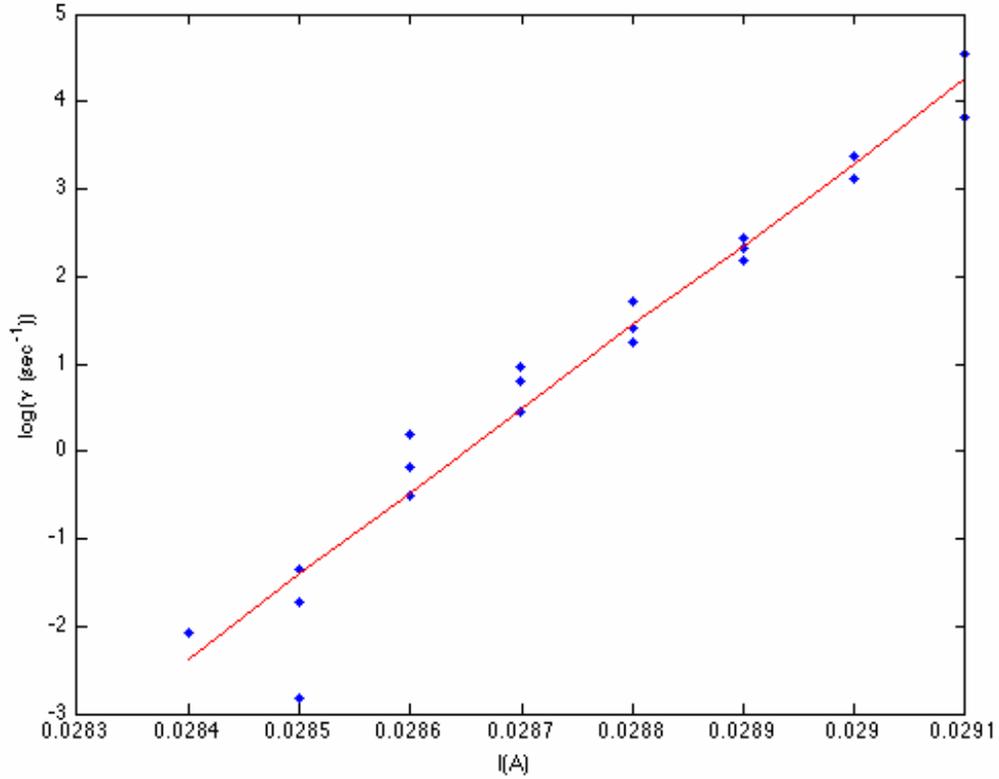

Figure S1. The solid points are the experimental telegraph noise frequencies for the asymmetric 24° grain boundary sample at T=10K. The line is a best fit to a quantum tunneling model as described in the text.